\documentclass[a4paper,11pt]{article}
\pdfoutput=1 

\usepackage{jheppub} 

\usepackage[T1]{fontenc} 
\usepackage{amsmath}
\usepackage{amsfonts}
\usepackage{amssymb}
\usepackage{epsfig,graphicx,bm}

\title{\boldmath Braids, normed division algebras, and Standard Model symmetries}

\author{N. Gresnigt}
\affiliation{Xi'an Jiaotong-Liverpool University,\\111 Ren'ai Road, Suzhou SIP, Higher Education Town, 215123, Jiangsu province, China}

\emailAdd{niels.gresnigt@xjtlu.edu.cn}

\abstract{
This paper represents a first attempt at unifying two promising attempts to understand the origin of the internal symmetries of leptons and quarks. It is shown that each of the four normed division algebras over the reals admits a representation of a circular braid group. For the complex numbers and the quaternions, the represented circular braid groups are $B_2$ and $B_3^c$, precisely those used to construct leptons and quarks as framed braids in the Helon model of Bilson-Thompson. It is then shown that these framed braids coincide with the states that span the minimal left ideals of the complex (chained) octonions, shown by Furey to describe one generation of leptons and quarks with unbroken $SU(3)_{c}$ and $U(1)_{em}$ symmetry. 

The identification of basis states of minimal ideals with certain framed braids is possible because the braiding in $B_2$ and $B_3^c$ in the Helon model are interchangeable. It is shown that the framed braids in the Helon model can be written as pure braid words in $B_3^c$ with trivial braiding in $B_2$, something which is not possible for framed braids in general.

}

\begin{document} 
\maketitle
\flushbottom

\section{Introduction}

In 2005, Bilson-Thompson proposed the Helon model in which one generation of leptons and quarks are identified as braidings of three ribbons with two crossings connected at the top and bottom via a node \cite{Bilson-Thompson2005}. These framed braids, with the additional structure that each ribbon can be twisted clockwise or anticlockwise by $2\pi$ (interpreted physically as electric charge), and satisfying certain conditions, map precisely to the first generation fermions of the Standard Model (SM). The original model has since been expanded into a complete scheme for the identification of the SM fermions and weak vector bosons for an unlimited series of generations \cite{Bilson-Thompson2009,Bilson-Thompson2008,Bilson-Thompson2012}. This topological model of elementary matter fits naturally into the context of Loop Quantum Gravity (LQG) which uses spin network graphs with edges labelled by representations of $SU(2)$. Instead labelling the edges by representations of the quantum group $SU_q(2)$ introduces a nonzero cosmological constant and requires that the edges be thickened to ribbons \cite{Bilson-Thompson2007}.

Furey in 2015 proposed an alternative explanation for the observed spacetime and internal symmetries of leptons and quarks in terms normed division algebras (NDAs) over the real numbers acting on themselves \cite{furey2016standard}. Her work builds on the initial results involving the octonions and particle physics by Gunaydin and Gursey in the 1970s \cite{gunaydin1973quark,gunaydin1974quark}. Complementary studies that look at the connection between NDAs and particle physics can be found in \cite{dixon1990derivation,dixon2004division,dixon2010division,manogue2010octonions}. The minimal ideals of the complex quaternions $\mathbb{C}\otimes\mathbb{H}$ are shown to contain exactly those representations of the Lorentz group corresponding to SM fermions. Furthermore, a Witt decomposition of the complex octonions $\mathbb{C}\otimes\overleftarrow{\mathbb{O}}$ (notation from section (\ref{chainedoctonions}))is shown to decompose the algebra into ideals whose basis states transform as a single generation of leptons and quarks under the unbroken unitary symmetries $SU(3)_c$ and $U(1)_{em}$. Additional work suggests that the same approach can give rise to exactly three generations \cite{furey2014generations}, and, at least for the case of leptons, automatically account for parity being maximally violated in weak interactions \cite{furey2018demonstration}. A similar model that merges both the spacetime and internal symmetries into a single copy of the complex Clifford algebra $C\ell(6)$ has been proposed by Stoica in 2017 \cite{stoica2017standard}. 

This paper makes a first attempt at unifying the model of Bilson-Thompson based on framed braids with the model of Furey based on minimal ideals of NDAs. In 2016, Kauffman and Lomonaco showed that Clifford algebras contain representations of circular braid groups and highlighted the close connection between the quaternions and topology, and how braiding is fundamental to the structure of fermionic physics \cite{kauffman2016braiding}. In the first part of this paper their work is extended, and isomorphisms between Clifford algebras and NDAs are used to show that each of the four NDAs contains a representation of a particular braid group. It is found that the complex numbers and quaternions contain representations of $B_2$ and $B_3^c$ respectively. These are precisely those braid groups from which the Helon model is constructed \cite{gresnigt2017quantum}.

Encouraged by this result, in the second part of this paper it is shown that the basis states of the minimal ideals of the complex octonions may be identified with precisely those framed braids that compose the Helon model, thereby establishing a connection between these two independent models. This identification of the basis states of minimal ideals with framed braids is made possible as a result of the braiding in $B_2$ and $B_3^c$ in the Helon model being interchangeable. It is shown that the $\pm 2\pi$ twists on ribbons (representing electric charge) can be written as products of certain braids in $B_3^c$ instead. These braids are then identified with the ladder operators  from which the minimal ideals of the complex octonions $\mathbb{C}\otimes\overleftarrow{\mathbb{O}}$ are constructed.

Following a review of the Helon model in section \ref{helonmodel} and the NDAs in section \ref{nda}, we then find the braid group representations admitted by each of the four NDAs in section \ref{cbt}. In section \ref{connection} we demonstrate that the braid groups represented by the NDAs are precisely those that appear in the Helon model. In section \ref{braidsasideals} it is shown that the basis states that span the minimal left ideals of the complex octonions coincide with the Helon braids.


\section{The Helon model}\label{helonmodelsection}

In this section we give the briefest of overviews of the Helon model, sufficient for our purposes. The reader is directed to the original paper in which the model was first presented \cite{Bilson-Thompson2005} for an in-depth presentation. 

\begin{figure}[h!]\label{helonmodel}
\centering
\includegraphics[scale=0.30]{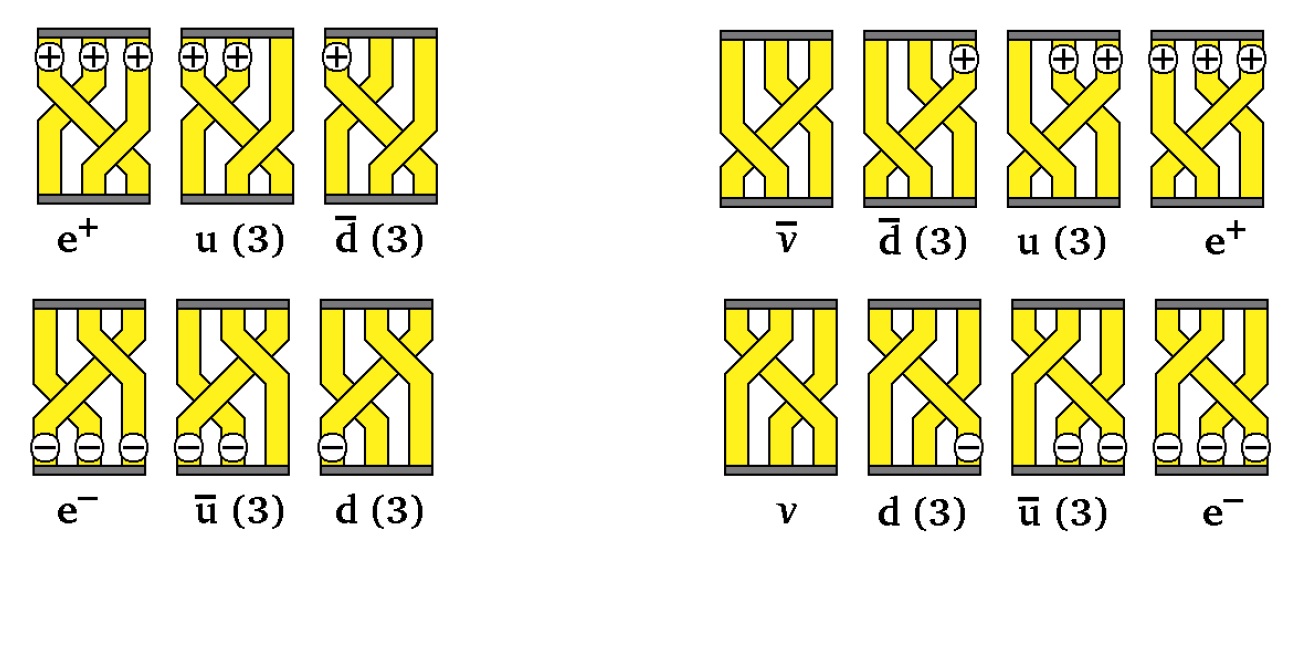}
\caption{The Helon model of Bilson-Thompson in which the first generation SM fermions are represented as braids of three (possibly twisted) ribbons. Used with permission. Source, \cite{Bilson-Thompson2005}.}
\end{figure}

The Helon model of Bilson-Thompson maps the simplest non-trivial braids consisting of three (twisted) ribbons and two crossings to the first generation of SM fermions. Quantized electric charges of particles are represented by integral twists of the ribbons of the braids, with a twist of $\pm 2\pi$ representing an electric charge of $\pm e/3$. The twist carrying ribbons, called Helons, are combined into triplets by connecting the tops of three ribbons to each other and likewise for the bottoms of the ribbons. The color charges of quarks and gluons are accounted for by the permutations of twists on certain braids, and simple topological processes are identified with the electroweak interaction, the color interaction, and conservation laws. The lack of twist on the neutrino braids means they only come in one handedness. The representation of first generation SM fermions in terms of braids is shown in Figure \ref{helonmodel}.

These braided structures may be embedded within a larger network of braided ribbons. Such a ribbon network is a generalization of a spin network, fundamental in LQG. The embedding of framed braids into ribbon networks make it possible to develop a unified theory of matter and spacetime in which both are emergent from the ribbon networks \cite{Bilson-Thompson2007}. 

\begin{figure}[h!]\label{embedding}
\centering
   \includegraphics[scale=0.3]{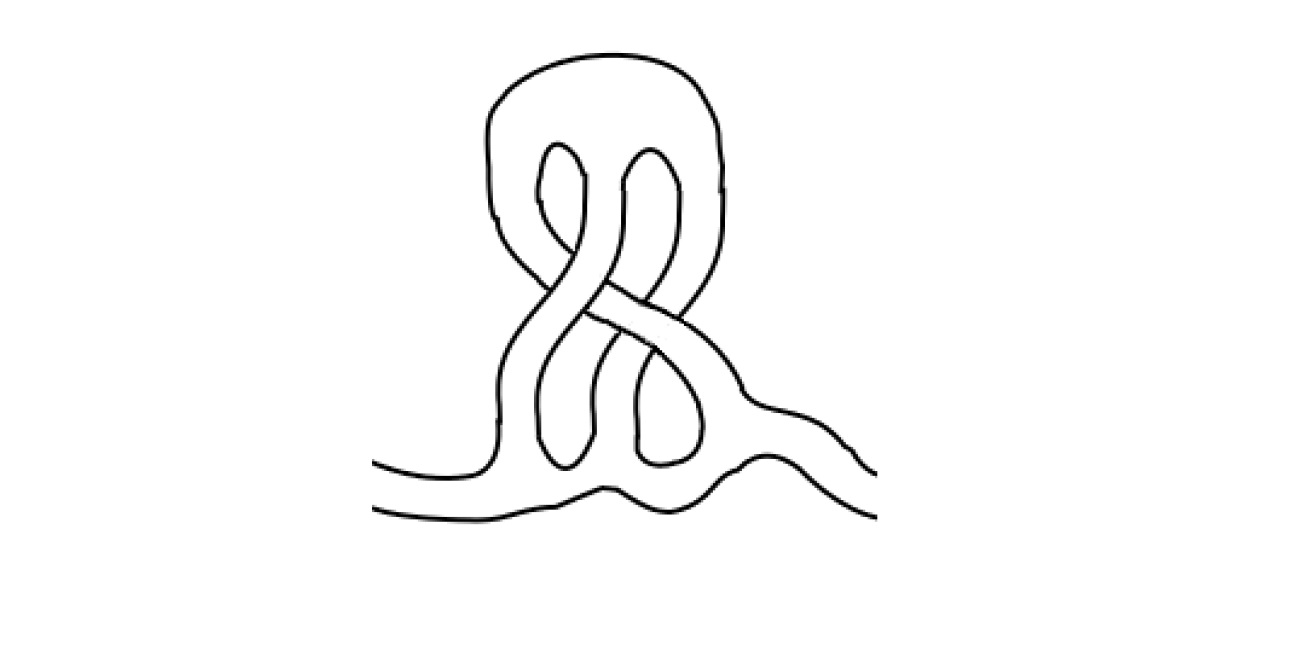}
\caption{The embedding of a Helon model braid into a ribbon network. Used with permission. Source, \cite{Bilson-Thompson2007}.}
\end{figure}

Within ribbon networks, these braided structures correspond to local noiseless subsystems which have been shown to exist in background independent theories where the microscopic quantum states are defined in terms of the embedding of a framed, or ribbon, graph in a three manifold and in which the allowed evolution moves are the standard local exchange and expansion moves (Pachner moves). Such noiseless subsystems are given by braided sets of $n$ edges joined at both ends by a set of connected nodes. The embedding into a ribbon network is possible by connecting (at least) one of the nodes to the rest of the ribbon graph. What the Helon model shows is that the simplest emergent local structures of such theories, when $n=3$, match precisely the first generation leptons and quarks. The embedding of a framed braid into a ribbon network is shown in Figure \ref{embedding}. 

Discrete symmetries have already been studied in the Helon model and may be defined on the braid in such a way that performing all three in any order leaves the braid unchanged \cite{he2008c}. Dynamics and interactions of braids have been studied in terms of evolution moves on trivalent and tetravalent spin network. Smolin and Wan \cite{smolin2008propagation} have shown that braid interactions in tetravalent such spin networks are understood in terms of dual Pachner moves. 

\subsection{The Artin braid groups}

The Artin braid group on $n$ strands is denoted by $B_n$ and is generated by elementary braids $\left\lbrace \sigma_1,...,\sigma_{n-1}\right\rbrace$ subject to the relations
\begin{eqnarray}\label{braidrelations}
\sigma_i \sigma_j=\sigma_j\sigma_i,\;\textrm{whenever}\; \vert i-j \vert > 1,\\
\sigma_i\sigma_{i+1}\sigma_i=\sigma_{i+1}\sigma_i \sigma_{i+1},\;\textrm{for}\; i=1,....,n-2.\label{braidrelations2}
\end{eqnarray}
The braid groups $B_n$ are an extension of the symmetric groups $S_n$ with the condition that the square of each generator being equal to one lifted.

In the framed braid group, each strand is thickened to a ribbon with the additional structure that now each ribbon may be twisted. Thus in addition to the braid generators $\sigma_1,...,\sigma_{n-1}$ of $B_n$, the framed braid group has the additional twist operators $t_1,...,t_n$. The framed braid group of $B_n$ is then defined by relations (\ref{braidrelations}, \ref{braidrelations2}) and additional relations
\begin{eqnarray}
t_it_j &=&t_jt_i,\;\textrm{for}\;\textrm{all}\;i,j,\\
\sigma_it_j&=&t_{\sigma_i(j)}\sigma_i,
\end{eqnarray}
where $\sigma_i(j)$ denotes the permutation induced on $(j)$ by $\sigma_i$. For example $\sigma_1(2)=(1)$ and $\sigma_1(3)=(3)$.

Finally, the inverse of a braid is its vertical reflection. This is an anti-automorphism so that, for example, $(\sigma_3\sigma_1\sigma_2^{-1})^{-1}=\sigma_2\sigma_1^{-1}\sigma_3^{-1}$.

\subsection{Semi-direct product structure of the Helon model}
In general, the twisting of the ribbons and the braiding of ribbons is not commutative, with the braidings inducing a permutation on the twists of the ribbons. The mathematical structure of framed braids is therefore that of the semi-direct product $B_3^c \ltimes (B_2)^3$. Because $B_2 \cong \frac{1}{2}\mathbb{Z}$ this can be rewritten as $(\frac{1}{2}\mathbb{Z})^3 \rtimes B_3^c=(\frac{1}{2}\mathbb{Z}\times \frac{1}{2}\mathbb{Z}\times \frac{1}{2}\mathbb{Z})\rtimes B_3^c$. An element of $(\frac{1}{2}\mathbb{Z}\times \frac{1}{2}\mathbb{Z}\times\frac{1}{2}\mathbb{Z})$ is denoted by a vector $[a_1,a_2,a_3]$ of multiples of half integers as in \cite{Bilson-Thompson2009,Bilson-Thompson2008}. A general framed braid may then be written in standard from with the twisting first followed by the braiding, as $([a_1,a_2,a_3],\Lambda)$\footnote{We will often simply write $[a_1,a_2,a_3]\Lambda$, dropping the parenthesis and comma.} where $[a_1,a_2,a_3]\in (\frac{1}{2}\mathbb{Z})^3$ is the twist word and $\Lambda\in B^c_3$ is the braid word.

Two framed braids may be multiplied together by first joining the bottom of the ribbons of the first braid to the tops of the ribbon of the second braid and then sliding (isotop) the twists from each component braid upward. Doing so, the twists carried by the first braid will get permuted by the second braid. The composition law may be written as
\begin{eqnarray}
\nonumber &{}&([a_1,a_2,a_3],\Lambda_1)([b_1,b_2,b_3],\Lambda_2)\\
\nonumber&=&(P_{\Lambda_1}([b_1,b_2,b_3])+[a_1,a_2,a_3],\Lambda_1\Lambda_2),\\
\nonumber&=&([b_{\pi(\Lambda_2)(1)},b_{\pi(\Lambda_2)(2)},b_{\pi(\Lambda_2)(3)}]+[a_1,a_2,a_3],\Lambda_1\Lambda_2),\\
&=&([b_{\pi(\Lambda_2)(1)}+a_1, b_{\pi(\Lambda_2)(2)}+a_2, b_{\pi(\Lambda_2)(3)}+a_3],\Lambda_1\Lambda_2)
\end{eqnarray}
where $\Lambda_1$ and $\Lambda_2$ are two braid words, $P_{\Lambda_i}$ is the permutation induced on $[a,b,c]$ by the braid word $\Lambda$, and $\pi:B_3^c\rightarrow S_3$ with $\pi(\sigma_1)=(12)$, $\pi(\sigma_2)=(23)$, $\pi(\sigma_3)=(31)$. One could instead slide the twists to the bottom of the braid, thus writing $(\Lambda, [a_1,a_2,a_3])$ but care must be taken to modify the composition law above respectively. Unless otherwise stated, the standard form will be considered to be the twist vector written first followed by the braiding.

As an example, consider the $u(3)$ up quark in the Helon model, as depicted in Figure \ref{helonmodel}. With the positive charges written at the top of the braid this can be written using the current notation as $([0,1,1],\sigma_2^{-1}\sigma_1)$. Similarly, one can write the anti up quark $\bar{u}(3)$ with the negative charges written at the bottom of the braid as $(\sigma_1^{-1}\sigma_2, [0,-1,-1])$. To write this in the standard from with the twisting first followed by the braiding (as for the example of the up quark) we can slide the charges along the ribbons. In the process they get permuted by the braiding, and one finds that
\begin{eqnarray}
(\sigma_1^{-1}\sigma_2, [0,-1,-1])=([-1,0,-1],\sigma_1^{-1}\sigma_2).
\end{eqnarray}

\section{Normed division algebras and Clifford algebras}\label{nda}

\subsection{Normed division algebras}
A division algebra is an algebra over a field where division is always possible, with the exception of division by zero. A normed division algebra (NDA) is a division algebra where in addition $\vert ab\vert=\vert a\vert\vert b\vert$\footnote{More precisely, a division algebra is a vector space over a field (in our case we are considering the field $\mathbb{R}$) which is also a ring with an identity under multiplication and in which $ax=b$ can be solved uniquely for $x$ unless $a=0$. A normed division algebra is also an integral domain, which means a ring in which $ab=0$ implies that $a=0$ or $b=0$.}. Nature admits only four NDAs over the reals: the real numbers $\mathbb{R}$, the complex numbers $\mathbb{C}$, the quaternions $\mathbb{H}$, and the octonions $\mathbb{O}$. Starting from the real numbers and generalizing to the complex numbers, one has to give up the ordered property of the reals. Generalizing in turn to the quaternions one furthermore gives up the commutativity of the reals and complex numbers. The quaternions are spanned by ${1,I,J,K}$ with $1$ being the identity and $I,J,K$ satisfying
\begin{eqnarray}
I^2=J^2=K^2=IJK=-1.
\end{eqnarray}

Finally, in moving to the octonions one has to give up the associativity of the reals, complex numbers, and quaternions. The lack of associativity of the octonions means their applications to physics have not been studied in as much details as for the other NDAs. An excellent introduction to the octonions, and in particular their relation to Clifford algebras, is given by Baez \cite{baez2002octonions}. The octonions are spanned by the identity $1=e_0$ and seven $e_i$ satisfying
\begin{eqnarray}
e_ie_j=-\delta_{ij}e_0+\epsilon_{ijk}e_k,
\end{eqnarray}
where 
\begin{eqnarray}
e_ie_0=e_0e_i=e_i,\;e_0^2=e_0,
\end{eqnarray}
and $\epsilon_{ijk}$ is a completely antisymmetric tensor with value +1 when $ijk = 123,\;145,\;176,\;246$, $257,\;347,\;365$. The multiplication of quaternions and octonions is shown in Figure 3.
\begin{figure}
\centering
\includegraphics[width=0.30\linewidth]{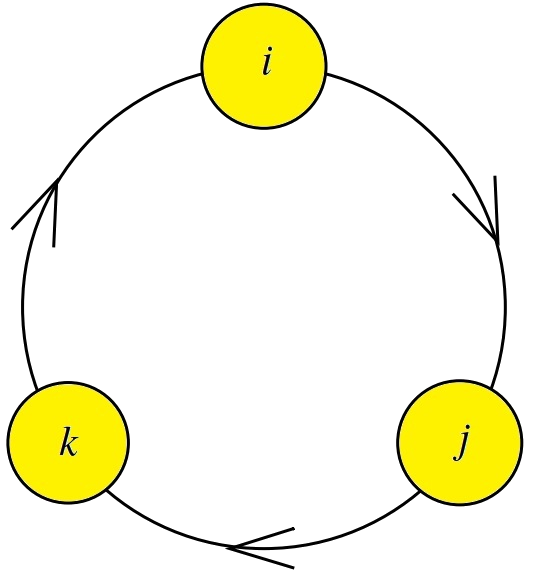}\label{NDA}
\qquad\qquad\qquad
\includegraphics[width=0.35\linewidth]{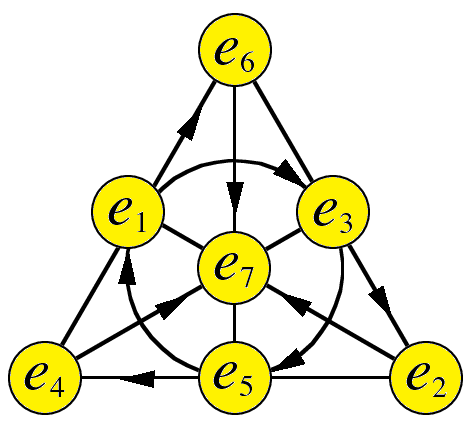}\label{NDA2}
\caption{Quaternion multiplication $I^2=J^2=K^2=IJK=-1$, and octonion multiplication represented using a Fano plane.}
\end{figure}

Every straight line in the Fano plane of the octonions (taken together with the identity) generates a copy of the quaternions, for example $\left\lbrace 1,e_4,e_1,e_6\right\rbrace $. The circle $\left\lbrace 1,e_1,e_3,e_5\right\rbrace $ also gives a copy of the quaternions, making for a total of seven copies of the quaternions embedded within the octonions.

Despite their non-associativity, the octonions have received some interest in attemps to describe the origin of quark and leptop structure and symmetry \cite{gunaydin1973quark,gunaydin1974quark,manogue2010octonions}. The automorphism group of the octonions is the exceptional Lie group $G_2$ which contains the physically important subgroups $SU(3)$ and $SU(2)\times SU(2)$. This earlier work has recently been complemented and extended by Furey \cite{furey2016standard,furey2014generations} and also Stoica \cite{stoica2017standard}.

\subsection{Clifford algebras}

Clifford algebras are the result of an attempt by William Clifford in 1876 to generalize the quaternions to higher dimensions and since then they have found many applications in physics \cite{doran2003gap,hestenes1966sta,Gresnigt2009,gresnigt2008,gresnigt2007sph}. They appear whenever spinors do, suggesting they likely play an important role in describing SM fermions. 

A real Clifford algebra on the vector space $\mathbb{R}^{p,q}$ equipped with a degenerate quadratic form is defined as the associative algebra generated by $p+q$ orthonormal basis elements $e_i$ satisfying
\begin{eqnarray}
e_ie_j&=&-e_je_i,\;\mathrm{for}\;i\neq j,\\
e_i^2&=&+1,\; 1\leq i\leq p,\\
e_i^2&=&-1,\; p<i\leq p+q.
\end{eqnarray}

One may likewise define complex Clifford algebras over complex spaces $\mathbb{C}^n$, denoted by $C\ell(n)$. However, doing so forfeits the signature and thus by extension much of the underlying geometry. For this reason we restrict ourselves whenever possible to real Clifford algebras over $\mathbb{R}^{p,q}$ which we write as $C\ell(p,q)$. The pair $(p,q)$ is called the signature of the underlying quadratic form.

A Clifford algebra $C\ell(p,q)$ has $2^{(p+q)}$ elements of different grades. We can write a general multivector $M\in C\ell(p,q)$ as
\begin{eqnarray}
M=\langle M \rangle_0+\langle M \rangle_1+\langle M \rangle_2+...+\langle M \rangle_{(p+q)},
\end{eqnarray}
where $\langle M \rangle_n$ contain the grade $n$ basis elements that are a product of $n$ distinct basis vectors $e_i$. 

The even elements of a Clifford algebra, those elements obtained from the Clifford product of an even number of basis elements form a subalgebra which is denoted $C\ell^+(p,q)$. There exists an isomorphism between  $C\ell^+(p,0)$ and $C\ell(0,p-1)$\footnote{More generally $Cl^+(p,q)\cong Cl(q,p-1)$ when $p>0$.} which we will make use of in the section. Explicitly, this isomorphism is given by
\begin{eqnarray}
\phi:C\ell(0,p-1)\rightarrow C\ell^+(p,0),\\
\phi(e_i)=e_ie_p,\qquad 1\leq i\leq p-1.
\end{eqnarray}
There also exist well-known isomorphisms between the associative NDAs and Clifford algebras. These are
\begin{eqnarray} 
\nonumber Cl(0,0)\cong\mathbb{R},\qquad Cl(0,1)\cong\mathbb{C}, \qquad  Cl(0,2) \cong \mathbb{H}.
\end{eqnarray}
Furthermore, the chained octonions $\overleftarrow{\mathbb{O}}$ defined below are isomorphic to $C\ell(0,6)$,
\begin{eqnarray}
C\ell(0,6)\cong\overleftarrow{\mathbb{O}}.
\end{eqnarray}

The matrix representations of real Clifford algebras up to $p+q=8$ as well as complex Clifford algebras can be found in Lounesto \cite{lounesto2001caa}. The matrix representations of larger Clifford algebras can be found using
\begin{eqnarray}
C\ell(p,q+8)&\cong& C\ell(p,q)\otimes \mathrm{Mat}(16,\mathbb{R}),\\
C\ell(p+8,q)&\cong& C\ell(p,q)\otimes \mathrm{Mat}(16,\mathbb{R}),
\end{eqnarray}
that is, $16\times 16$ matrices with entries in $C\ell(p,q)$. Some of the matrix representations relating to the larger Clifford algebras are
\begin{eqnarray}
\nonumber C\ell(0,6)\cong \mathbb{R}(8),\qquad C\ell(6,0)\cong \mathbb{H}(4),\\
\nonumber C\ell(0,7)\cong {}^2\mathbb{R}(8),\qquad C\ell(7,0)\cong \mathbb{C}(8),
\end{eqnarray}
and for complex Clifford algebras
\begin{eqnarray}
\nonumber C\ell(6)\cong \mathbb{C}(8),\qquad C\ell(7)\cong {}^2\mathbb{C}(8).
\end{eqnarray}
Important in what follows is the isomorphism
\begin{eqnarray}
\nonumber C\ell(6)\cong \mathbb{C}\otimes\overleftarrow{\mathbb{O}}.
\end{eqnarray}

Finally, there are three important involutions. These are defined as follows
\begin{eqnarray}
\hat{u}&:&e_i\mapsto -e_i \qquad \textrm{grade\;involution},\\
\tilde{u}&:&e_i....e_n \mapsto e_n....e_i\qquad \textrm{reversion},\\
\bar{u}&:& e_i...e_n \mapsto (-e_n)....(-e_i) \qquad \textrm{Clifford\;conjugation}, 
\end{eqnarray}
where $u$ is general element of $C\ell(p,q)$. Whereas grade involution is an automorphism, both reversion and Clifford conjugation are anti-automorphisms. The effects of these involutions on the multivectors of, for example, $C\ell(0,3)$ are
\begin{eqnarray}
\hat{u}&=&\langle u \rangle_0-\langle u \rangle_1+\langle u \rangle_2-\langle u \rangle_3\qquad \textrm{grade\;involution},\\
\tilde{u}&=&\langle u \rangle_0+\langle u \rangle_1-\langle u \rangle_2-\langle u \rangle_3\qquad \textrm{reversion},\\
\bar{u}&=&\langle u \rangle_0-\langle u \rangle_1-\langle u \rangle_2+\langle u \rangle_3\qquad \textrm{Clifford\;conjugation}
\end{eqnarray}

\subsection{Minimal left ideals of the complex chained octonions}\label{ideals}

Two main results of Furey's thesis are that the generalized ideals of the complex quaternions
describe consicely all of the Lorentz group representations found in the SM and that the minimal left ideals of the complex octonions mirror the behaviour of a single generation of leptons and quarks with unbroken $SU(3)_c$ and $U(1)_{em}$ symmetry \cite{furey2016standard}. Relevant to what is to follow are the minimal left ideals of the complex octonions, and for this reason we review the construction of these ideals briefly. A more detailed construction of minimal left ideals in general, and specifically for the case of the complex octonions can be found in sections 4.5 and 6.6 of the above cited work.

The ideals are constructed using the Witt decomposition for (complex) $C\ell(6)$ which is isomorphic to $\mathbb{C}\otimes\overleftarrow{\mathbb{O}}$. The first ideal is written in terms of a primitive idempotent $\omega\omega^{\dagger}=\alpha_1\alpha_2\alpha_3\alpha_3^{\dagger}\alpha_2^{\dagger}\alpha_1^{\dagger}$ defined in terms of the basis vectors
\begin{eqnarray}
\alpha_1\equiv\frac{1}{2}(-e_5+ie_4),\qquad \alpha_2\equiv\frac{1}{2}(-e_3+ie_1),\qquad \alpha_3\equiv \frac{1}{2}(-e_6+ie_2).
\end{eqnarray}
These basis vectors satisfy the anticommutation relation
\begin{eqnarray}
\left\lbrace \alpha_i,\alpha_j \right\rbrace =0,
\end{eqnarray}
and can be identified with lowering operators. The hermitian conjugate simultaneously maps $i\mapsto -i$ and $e_i\mapsto -e_i$ so that
\begin{eqnarray}
\alpha_1^{\dagger}\equiv\frac{1}{2}(e_5+ie_4),\qquad \alpha_2^{\dagger}\equiv\frac{1}{2}(e_3+ie_1),\qquad \alpha_3^{\dagger}\equiv \frac{1}{2}(e_6+ie_2).
\end{eqnarray}
satisfying the anticommutation relations
\begin{eqnarray}
\left\lbrace \alpha_i^{\dagger},\alpha_j^{\dagger} \right\rbrace =0,\qquad \left\lbrace \alpha_i,\alpha_j^{\dagger} \right\rbrace =\delta_{ij}.
\end{eqnarray}

The first minimal left ideal is given by $S^u\equiv \mathbb{C}\otimes\overleftarrow{\mathbb{O}}\omega\omega^{\dagger}$
\begin{eqnarray}
\nonumber S^u\equiv &{}&\\
\nonumber &{}&\;\;\nu \omega\omega^{\dagger}+\\
\nonumber \bar{d}^r\alpha_1^{\dagger}\omega\omega^{\dagger} &+& \bar{d}^g\alpha_2^{\dagger}\omega\omega^{\dagger} + \bar{d}^b\alpha_3^{\dagger}\omega\omega^{\dagger}\\
\nonumber u^r\alpha_3^{\dagger}\alpha_2^{\dagger}\omega\omega^{\dagger} &+& u^g\alpha_1^{\dagger}\alpha_3^{\dagger}\omega\omega^{\dagger} + u^b\alpha_2^{\dagger}\alpha_1^{\dagger}\omega\omega^{\dagger}\\
&+& e^{+}\alpha_3^{\dagger}\alpha_2^{\dagger}\alpha_1^{\dagger}\omega\omega^{\dagger},
\end{eqnarray}
where $\nu$, $\bar{d}^r$ etc. are suggestively labeled complex coefficients. The complex conjugate system analogously gives a second linearly independent minimal left ideal
\begin{eqnarray}
\nonumber S^d\equiv &{}&\\
\nonumber &{}&\;\;\bar{\nu} \omega^{\dagger}\omega+\\
\nonumber d^r\alpha_1\omega^{\dagger}\omega &+& d^g\alpha_2\omega^{\dagger}\omega + d^b\alpha_3\omega^{\dagger}\omega\\
\nonumber \bar{u}^r\alpha_3\alpha_2\omega^{\dagger}\omega &+& \bar{u}^g\alpha_1\alpha_3\omega^{\dagger}\omega + \bar{u}^b\alpha_2\alpha_1\omega^{\dagger}\omega\\
&+& e^{-}\alpha_3\alpha_2\alpha_1\omega^{\dagger}\omega,
\end{eqnarray}

It can be shown that these representations of the minimal left ideals are invariant to the color and electromagnetic symmetries $SU(3)_c$ and $U(1)_{em}$ and each of the basis states in the ideals transforms as a specific lepton or quark under these symmetries as indicated by their suggestively labeled complex coefficients.  
\section{Normed division algebra representations of circular Artin braid groups}\label{cbt}

\subsection{The Clifford Braiding Theorem}
In 2016 Kauffman and Lomonaco showed, in what they call the Clifford Braiding Theorem (CBT), that Clifford algebras contain representations of (circular) braid groups \cite{kauffman2016braiding}.

For a Clifford algebra $Cl(n,0)$ over the real numbers generated by linearly independent elements $\left\lbrace e_1,e_2,...,e_n\right\rbrace $ with $e_k^2=1$ for all $k$ and $e_ke_l=-e_le_k$ for $k\neq l$, the algebra elements $\sigma_k=\frac{1}{\sqrt{2}}(1+e_{k+1}e_k)$ form a representation of the circular\footnote{A circular braid on $n$ strings has $n$ strings attached to the outer edges of two circles which lie in parallel planes in $R^3$.} Artin braid group $B_n$. This means that the set of braid generators $\left\lbrace  \sigma_1,\sigma_2,...,\sigma_n\right\rbrace $ where
\begin{eqnarray}
\sigma_k &=& \frac{1}{\sqrt{2}}(1+e_{k+1}e_k),\;\textrm{whenever}\; 1\leq k <n,\\
\sigma_n &=&\frac{1}{\sqrt{2}}(1+e_1e_n),
\end{eqnarray}
satisfy the braid relations (\ref{braidrelations}). An important point is that the order of the braid generators represented this way is eight. Although the original theorem as found in \cite{kauffman2016braiding} assumes that $e_k^2=1$ for all $k$, the proof likewise holds when $e_k^2=-1$, as is easily checked. The important point is that it fails to hold for a general Clifford algebra $Cl(p,q)$ of mixed signature.

The braid generators are composed of the scalar and a subset of the bivectors elements of $Cl(n,0)$, and therefore live in the even subalgebra $Cl^+(n,0)\cong C\ell(0,n-1)$. In what follows we use the known isomorphisms between the NDAs and Clifford algebras listed earlier to determine which braid groups may be represented by the NDAs.

\subsection{A representation of the Artin braid group $B_2$ from $\mathbb{C}$}

The complex numbers $\mathbb{C}$ with basis $\left\lbrace 1, i\right\rbrace$ are isomorphic to the Clifford algebra $Cl(0,1)$ with $e_1^2=-1$. Given this isomorphism it means one also has an isomorphism with $Cl^+(2,0)$, the even part of $Cl(2,0)$. Therefore, the complex number algebra $\mathbb{C}$ admits a representation of the braid group $B_2$. In this case the Artin braid group is equivalent to the circular Artin braid group $B^c_2\cong B_2$. The single braid generator $\sigma_1$ can be represented in terms of the scalar and bivector of $Cl(2,0)$, so that
\begin{eqnarray}
\sigma_1=\frac{1}{\sqrt{2}}(1+e_2e_1),\qquad \sigma_1^{-1}=\frac{1}{\sqrt{2}}(1-e_2e_1),
\end{eqnarray} 
with the inverse generators defined by inserting a minus sign in front of the bivector terms. Alternatively, in $C\ell(0,1)$ the braid generator and its inverse take the form
\begin{eqnarray}
\sigma_1=\frac{1}{\sqrt{2}}(1-e_1),\qquad \sigma_1^{-1}=\frac{1}{\sqrt{2}}(1+e_1).
\end{eqnarray}
Using the isomorphism $Cl^{+}(2,0)\cong Cl(0,1)\cong\mathbb{C}$, $\mathbb{C}$ gives a representation of $B_2$ with the braid generator expressed as
\begin{eqnarray}
\sigma_1=\frac{1}{\sqrt{2}}(1+i),\qquad \sigma_1^{-1}=\frac{1}{\sqrt{2}}(1-i).
\end{eqnarray}
The map from $\sigma_1\mapsto \sigma_1^{-1}$ can be seen as complex conjugation $*:i\mapsto -i$ in $\mathbb{C}$, as reversion in $C\ell(2,0)$, and Clifford conjugation (or alternatively grade involution) in $C\ell(0,1)$.

The order of $\sigma_1$ is eight, and one can readily check that
\begin{eqnarray}
\nonumber \sigma_1&=&(1/\sqrt{2})(1+i)=\sigma_1^{-7},\qquad \sigma_1^2=i=\sigma_1^{-6},\qquad \sigma_1^3=-(1/\sqrt{2})(1-i)=\sigma_1^{-5},\\
\nonumber \sigma_1^4&=&-1=\sigma_1^{-4},\qquad \sigma_1^5=-(1/\sqrt{2})(1+i)=\sigma_1^{-3},\qquad \sigma_1^6=-i=\sigma_1^{-2},\\
\nonumber \sigma_1^7&=&(1/\sqrt{2})(1-i)=\sigma_1^{-1},\qquad \sigma_1^8=1=\sigma_1^{-8},\qquad \sigma_1^9=(1/\sqrt{2})(1+i)=\sigma_1.
\end{eqnarray}
This means is that not every braid in $B_2$ can be represented in terms of $\mathbb{C}$. 

\subsection{A representation of the circular Artin braid group $B^c_3$ from $\mathbb{H}$}

Moving on to the quaternions $\mathbb{H}$, one can use the isomorphism $\mathbb{H}\cong Cl(0,2)\cong Cl^+(3,0)$ to find a quaternionic representation of the braid group $B^c_3$. $Cl(0,2)$ is spanned by $\lbrace 1,e_1,e_2,e_1e_2=e_{12}\rbrace$ with
\begin{eqnarray}
e_1^2=e_2^2=e_{12}^2&=&e_1e_2e_{12}=-1,\\
e_1e_2=-e_2e_1,\; e_1e_{12}&=&-e_{12}e_1,\; e_2e_{12}=-e_{12}e_2.
\end{eqnarray}
One can thus identify $e_1=I,\;e_2=J,\;e_{12}=K$ to obtain a copy of $\mathbb{H}$. The even subalgebra of   $Cl(3,0)$ contains three bivectors (and the scalar) which may be related to braid generators for $B^c_3$
\begin{eqnarray}
\sigma_1=\frac{1}{\sqrt{2}}(1+e_2e_1),\; \sigma_2=\frac{1}{\sqrt{2}}(1+e_3e_2),\; \sigma_3=\frac{1}{\sqrt{2}}(1+e_1e_3).
\end{eqnarray}
It is readily checked that
\begin{eqnarray}
\sigma_1\sigma_2\sigma_1=\sigma_2\sigma_1\sigma_2,\;\;
\sigma_2\sigma_3\sigma_2=\sigma_3\sigma_2\sigma_3,\;\;
\sigma_3\sigma_1\sigma_3=\sigma_1\sigma_3\sigma_1.
\end{eqnarray}
The representation in terms of $C\ell(0,2)$ is given in the Appendix. In terms of the quaternions we have
\begin{eqnarray}
\nonumber\sigma_1&=\frac{1}{\sqrt{2}}(1+I),\;\;\sigma_2=\frac{1}{\sqrt{2}}(1+J),\;\;\sigma_3=\frac{1}{\sqrt{2}}(1+K),
\end{eqnarray}
with the inverses again obtained by inserting a minus sign, corresponding to taking the quaternion conjugate which maps $I,J,K$ to their negatives. In $C\ell^+(3,0)$ and $C\ell(0,2)$, the inverse braid generators are again obtained via reversion and Clifford conjugation respectively.

\subsection{A representation of the circular Artin braid group $B^c_7$ from $\mathbb{O}$}\label{chainedoctonions}

In addition to being non-commutative, the octonion algebra $\mathbb{O}$ is also non-associative making a matrix representation of the algebra impossible. However, by defining a standard ordering to any product of octonions, it is possible to recover an associative description of the octonions. Let $n$, $m$, $p$, and $f$ be four octonions. One defines the octonion chain $\overleftarrow{pnm}$ as the map $\overleftarrow{pnm}:f\mapsto p(n(mf))$. One can generalize this to a product of arbitrary many octonions. The resulting algebra is the chained octonions $\overleftarrow{\mathbb{O}}$ which is associative and can be shown to be isomorphic to $Cl(0,6)$ \cite{furey2016standard}. 

Using the isomorphism $\overleftarrow{\mathbb{O}}\cong Cl(0,6)\cong Cl^+(7,0)$ one finds a representation of the braid group $B^c_7$ in terms of the chained octonions
\begin{eqnarray}
\sigma_i&=\frac{1}{\sqrt{2}}(1+\overleftarrow{e_{i+1}e_{i}}),\;\;\sigma_7=\frac{1}{\sqrt{2}}(1+\overleftarrow{e_1e_7}),
\end{eqnarray}
again with period eight. The explicit braid group representations in terms of $C\ell^+(7,0)$ and $C\ell(0,6)$ are given in the Appendix. Once again, reversion and Clifford conjugation maps the braid generators to their inverses for these two Clifford algebras respectively.

In summary, the NDAs provide the following (circular) braid group representations:
\begin{eqnarray}
\nonumber &\mathbb{C}&\cong Cl(0,1)\cong Cl^+(2,0)\rightarrow B_2,\\
\nonumber &\mathbb{H}&\cong Cl(0,2)\cong Cl^+(3,0)\rightarrow B^c_3,\\
\nonumber &\overleftarrow{\mathbb{O}}&\cong Cl(0,6)\cong Cl^+(7,0)\rightarrow B^c_7.
\end{eqnarray}

\subsection{Connecting the Helon model with normed division algebras}\label{connection}

The framed braids that represent fermions in the Helon model are constructed out of two braid groups, $B_2$ and $B_3$. The twisting of the ribbons, representing (quantised) electric charge corresponds to elements of $B_2$. When the ribbon is twisted the two edges of the ribbon braid one another. Additionally, the braiding of three ribbons forms a braid word in $B_3$. Furthermore, the individual ribbons of the braids are connected together at the top and bottom via a node. This arrangement where ribbons are connected at both ends is equivalent to two parallel disks connected by three ribbons. One therefore not only has $B_3$ but rather the circular braid group $B_3^c$.

An interesting observation is that these two braid groups are precisely those represented by the complex numbers $\mathbb{C}$ and the quaternions $\mathbb{H}$, suggesting it may be possible to connect the Helon model with the NDA model. Indeed this is not the only hint at a close connection between the two models and in the next section it is shown that by identifying the ladder operators $\alpha_i$ and $\alpha_i^{\dagger}$ with certain braids in $B_3^c$, the basis states of the minimal left ideals of the complex octonions become identical to the framed braids in the Helon model. This is the main result of this paper.

\section{Helon braids as basis states of minimal left ideals of $\mathbb{C}\otimes\mathbb{O}$}\label{braidsasideals}

\subsection{Interchanging between braiding and twisting}

It was demonstrated in \cite{Bilson-Thompson2009} that any braiding can always be exchanged for twisting (in the case for three ribbon braids). This means that any element $([a,b,c],\Lambda)\in (B_2)^3\rtimes B_3^c$ may always be rewritten as $[a',b',c']\in (B_2)^3$ in which the braiding in $B_3^c$ is trivial. The framed braids in the Helon model can therefore be written purely in terms of twist vectors. For example, in Figure 4, it is shown how the braiding induced by the generator $\sigma_1$ may be exchanged for twisting. 
\begin{figure}[h!]\label{braidtotwist2}
\centering
   \includegraphics[scale=0.15]{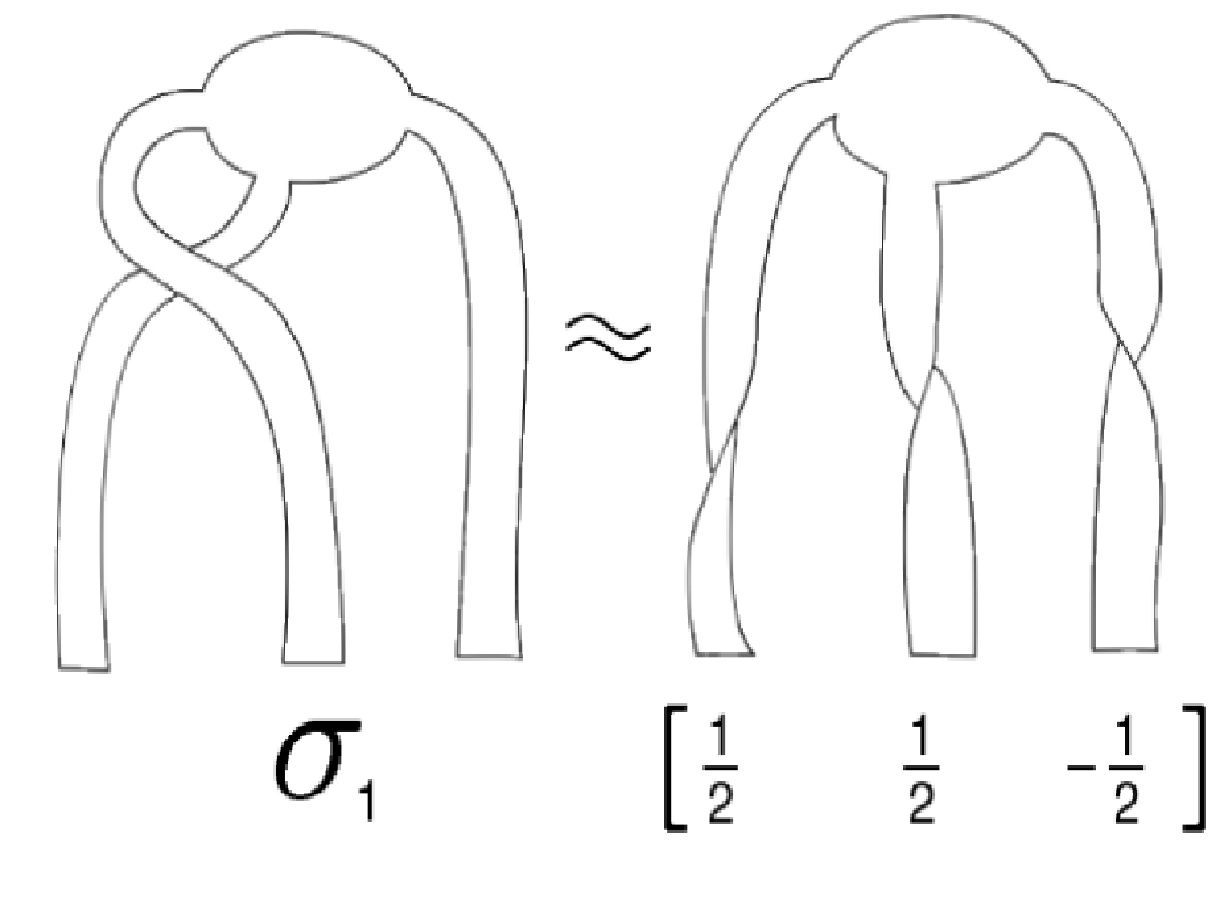}
\caption{The twisting and braiding in CRBs (consisting of two or three ribbons) is interchangeable. Source, \cite{Bilson-Thompson2007}.}
\end{figure}

The braid generators of the circular Artin braid group $B^c_3$ can be written as twist vectors as follows:
\begin{eqnarray}\label{braidgenerators}
\sigma_1\rightarrow \left[ \frac{1}{2},\frac{1}{2},-\frac{1}{2}\right] ,\qquad \sigma^{-1}_1\rightarrow \left[ -\frac{1}{2},-\frac{1}{2},\frac{1}{2}\right] ,\\
\sigma_2\rightarrow \left[ -\frac{1}{2},\frac{1}{2},\frac{1}{2}\right] ,\qquad\sigma^{-1}_2\rightarrow \left[ \frac{1}{2},-\frac{1}{2},-\frac{1}{2}\right],\\
\sigma_3\rightarrow \left[ \frac{1}{2},-\frac{1}{2},\frac{1}{2}\right] ,\qquad\sigma^{-1}_3\rightarrow \left[ -\frac{1}{2}, \frac{1}{2},-\frac{1}{2}\right] .
\end{eqnarray}
In turning a general braid into a pure twist vector one has to be careful to take into account the permutations induced by braidings. Thus, for example
\begin{eqnarray}
\nonumber[0,1,0]\sigma_1\sigma_2&=&\left(P_{\sigma_1}[0,1,0]+\left[\frac{1}{2},\frac{1}{2},-\frac{1}{2}\right]\right)\sigma_2,\\
\nonumber &=&\left([1,0,0]+\left[\frac{1}{2},\frac{1}{2},-\frac{1}{2}\right]\right)\sigma_2,\\
\nonumber &=&\left[\frac{3}{2},\frac{1}{2},-\frac{1}{2}\right]\sigma_2,\\
\nonumber &=&P_{\sigma_2}\left[\frac{3}{2},\frac{1}{2},-\frac{1}{2}\right]+\left[ -\frac{1}{2},\frac{1}{2},\frac{1}{2}\right],\\
\nonumber &=&\left[\frac{3}{2},-\frac{1}{2},\frac{1}{2}\right]+\left[ -\frac{1}{2},\frac{1}{2},\frac{1}{2}\right],\\
 &=&[1,0,1],
\end{eqnarray}
where by $P_{\sigma_i}[a,b,c]$ we denote the permutation on $[a,b,c]$ induced by the braiding $\sigma_i$. Unless otherwise stated the action is always from left to right.

One might instead want to go the other way, that is write a framed braid in pure braid form with trivial twisting $([0,0,0])$. This is in general not possible, but is possible for the particular braids in the Helon model. To see this, notice that the twists on an arbitrary Helon braid, ignoring the braiding for the time being, corresponds to one of the twist vectors $[\pm 1,0,0],\;[\pm 1,\pm 1,0],\;[\pm 1,\pm 1,\pm 1]$ and cyclic.

We leave it for the reader to verify that
\begin{eqnarray}\label{twists1}
\nonumber [0,0,0](\sigma_2\sigma_3)&=&[1,0,0],\\ 
\nonumber [0,0,0](\sigma_3\sigma_1)&=&[0,1,0],\\ 
\nonumber [0,0,0](\sigma_1\sigma_2)&=&[0,0,1],\\
\nonumber [0,0,0](\sigma_3\sigma_1)(\sigma_1\sigma_2)&=&[1,0,1],\\ 
\nonumber [0,0,0](\sigma_1\sigma_2)(\sigma_2\sigma_3)&=&[1,1,0],\\ 
\nonumber [0,0,0](\sigma_2\sigma_3)(\sigma_3\sigma_1)&=&[0,1,1],\\
\left[0,0,0 \right](\sigma_2\sigma_3)(\sigma_3\sigma_1)(\sigma_1\sigma_2)&=&[1,1,1],
\end{eqnarray}
and similarly
\begin{eqnarray}\label{twists2}
\nonumber[0,0,0](\sigma_2^{-1}\sigma_3^{-1})&=&[-1,0,0],\\
\nonumber[0,0,0](\sigma_3^{-1}\sigma_1^{-1})&=&[0,-1,0],\\ 
\nonumber[0,0,0](\sigma_1^{-1}\sigma_2^{-1})&=&[0,0,-1],\\ 
\nonumber[0,0,0](\sigma_3^{-1}\sigma_1^{-1})(\sigma_1^{-1}\sigma_2^{-1})&=&[-1,0,-1],\\
\nonumber[0,0,0](\sigma_1^{-1}\sigma_2^{-1})(\sigma_2^{-1}\sigma_3^{-1})&=&[-1,-1,0],\\
\nonumber[0,0,0](\sigma_2^{-1}\sigma_3^{-1})(\sigma_3^{-1}\sigma_1^{-1})&=&[0,-1,-1],\\ 
\left[ 0,0,0\right](\sigma_2^{-1}\sigma_3^{-1})(\sigma_3^{-1}\sigma_1^{-1})(\sigma_1^{-1}\sigma_2^{-1})&=&[-1,-1,- 1],
\end{eqnarray}
It should be noted that the representation of a twist vector in pure braid from is in general not unique. For example, $[0,0,0]\sigma_2\sigma_1=[1,0,0]$ also and $[0,0,0](\sigma_1^{-1}\sigma_3^{-1})(\sigma_3^{-1
}\sigma_2^{-1})(\sigma_2^{-1}\sigma_1^{-1})=[-1,-1,-1]$.

\subsection{Braid representations of minimal left ideals of the complex chained octonions}

If we now consider the neutrino in the Helon model, written as the braid $\sigma_2^{-1}\sigma_1$ and with no twisting of the ribbons, then the up quark, anti-down quark and positron can be considered excitations of the neutrino in the sense that their representations are obtained by adding twist to the ribbons that compose the neutrino but leaving the underlying braid structure unchanged. One can then write these fermions in braid-only form where the twisting has been removed, using eqn. (\ref{twists1}) and eqn. (\ref{twists2}) as
\begin{eqnarray}\label{positivestates}
\nonumber\nu &\rightarrow & [0,0,0](\sigma_2^{-1}\sigma_1)=(\sigma_2^{-1}\sigma_1),\\
\nonumber\bar{d}^r &\rightarrow & [0,0,1](\sigma_2^{-1}\sigma_1)=(\sigma_1\sigma_2)(\sigma_2^{-1}\sigma_1),\\
\nonumber\bar{d}^g &\rightarrow & [0,1,0](\sigma_2^{-1}\sigma_1)=(\sigma_3\sigma_1)(\sigma_2^{-1}\sigma_1),\\
\nonumber\bar{d}^b &\rightarrow & [1,0,0](\sigma_2^{-1}\sigma_1)=(\sigma_2\sigma_3)(\sigma_2^{-1}\sigma_1),\\
\nonumber u^r &\rightarrow & [0,1,1](\sigma_2^{-1}\sigma_1)=(\sigma_2\sigma_3)(\sigma_3\sigma_1)(\sigma_2^{-1}\sigma_1),\\
\nonumber u^g &\rightarrow & [1,1,0](\sigma_2^{-1}\sigma_1)=(\sigma_1\sigma_2)(\sigma_2\sigma_3)(\sigma_2^{-1}\sigma_1),\\
\nonumber u^b &\rightarrow & [1,0,1](\sigma_2^{-1}\sigma_1)=(\sigma_3\sigma_1)(\sigma_1\sigma_2)(\sigma_2^{-1}\sigma_1),\\
 e^+ &\rightarrow & [1,1,1](\sigma_2^{-1}\sigma_1)=(\sigma_2\sigma_3)(\sigma_3\sigma_1)(\sigma_1\sigma_2)(\sigma_2^{-1}\sigma_1).
\end{eqnarray}

The main result of this paper is that if one now identifies
\begin{eqnarray}\label{alphadagger}
(\sigma_1\sigma_2)=\alpha_1^{\dagger},\;\;(\sigma_3\sigma_1)=\alpha_2^{\dagger},\;\;(\sigma_2\sigma_3)=\alpha_3^{\dagger},
\end{eqnarray}
together with
\begin{eqnarray}
\sigma_2^{-1}\sigma_1=\omega\omega^{\dagger},
\end{eqnarray}
and substitutes into equation (\ref{negativestates}), then the minimal left ideal $S^u$ of the complex octonions (repeated below for convenience) is recovered
\begin{eqnarray}
\nonumber S^u\equiv &{}&\\
\nonumber &{}&\;\;\nu \omega\omega^{\dagger}+\\
\nonumber \bar{d}^r\alpha_1^{\dagger}\omega\omega^{\dagger} &+& \bar{d}^g\alpha_2^{\dagger}\omega\omega^{\dagger} + \bar{d}^b\alpha_3^{\dagger}\omega\omega^{\dagger}\\
\nonumber u^r\alpha_3^{\dagger}\alpha_2^{\dagger}\omega\omega^{\dagger} &+& u^g\alpha_1^{\dagger}\alpha_3^{\dagger}\omega\omega^{\dagger} + u^b\alpha_2^{\dagger}\alpha_1^{\dagger}\omega\omega^{\dagger}\\
&+& e^{+}\alpha_3^{\dagger}\alpha_2^{\dagger}\alpha_1^{\dagger}\omega\omega^{\dagger},
\end{eqnarray}
where the action of the basis states in the ideal is on the identity $[0,0,0]$ from left to right. Thus for example,
\begin{eqnarray}
\nonumber u^g &\rightarrow & [1,1,0](\sigma_2^{-1}\sigma_1)=[0,0,0](\sigma_1\sigma_2)(\sigma_2\sigma_3)(\sigma_2^{-1}\sigma_1)=[0,0,0]\alpha_1\alpha_3\omega^{\dagger}\omega.
\end{eqnarray}

Next consider the antiparticles, corresponding (in the Helon model) to the vertical reflections. The vertical reflection inverts both the braidings, and the signs of the twists as well and further moves the twists from the top of the braid to the bottom of the braid. This is evident from Figure \ref{helonmodel}. To illustrate consider the $u^b$ quark written as a pure braid word in Eq.(\ref{positivestates}) as $(\sigma_3\sigma_1)(\sigma_1\sigma_2)(\sigma_2^{-1}\sigma_1)$. It follows that for its antiparticle, the pure braid word must be
\begin{eqnarray}
\bar{u}^b\rightarrow (\sigma_1^{-1}\sigma_2)(\sigma_2^{-1}\sigma_1^{-1})(\sigma_1^{-1}\sigma_3^{-1}).
\end{eqnarray}
The last two terms in parenthesis are responsible for generating the twist vector but because of the vertical reflection the action is now from right to left. To be consistent this should be rewritten so that the action is from left to right to give
\begin{eqnarray}
\nonumber \bar{u}^b &\rightarrow& (\sigma_1^{-1}\sigma_2)(\sigma_2^{-1}\sigma_1^{-1})(\sigma_1^{-1}\sigma_3^{-1})[0,0,0],\\
\nonumber &=&(\sigma_1^{-1}\sigma_2)[0,0,0](\sigma_3^{-1}\sigma_1^{-1})(\sigma_1^{-1}\sigma_2^{-1}),\\
&=&(\sigma_1^{-1}\sigma_2)[-1,0,-1].\label{example}
\end{eqnarray}
Doing the same for the other antiparticles gives
\begin{eqnarray}\label{negativestates}
\nonumber \bar{\nu} &\rightarrow & (\sigma_1^{-1}\sigma_2)[0,0,0]=(\sigma_1^{-1}\sigma_2),\\
\nonumber d^r &\rightarrow & (\sigma_1^{-1}\sigma_2)[0,0,-1]=(\sigma_1^{-1}\sigma_2)(\sigma_1^{-1}\sigma_2^{-1}),\\
\nonumber d^g &\rightarrow & (\sigma_1^{-1}\sigma_2)[0,-1 ,0]=(\sigma_1^{-1}\sigma_2)(\sigma_3^{-1}\sigma_1^{-1}),\\
\nonumber d^b &\rightarrow & (\sigma_1^{-1}\sigma_2)[-1,0,0]=(\sigma_1^{-1}\sigma_2)(\sigma_2^{-1}\sigma_3^{-1}),\\
\nonumber \bar{u}^r &\rightarrow & (\sigma_1^{-1}\sigma_2)[0,-1,-1]=(\sigma_1^{-1}\sigma_2)(\sigma_2^{-1}\sigma_3^{-1})(\sigma_3^{-1}\sigma_1^{-1}),\\
\nonumber \bar{u}^g &\rightarrow & (\sigma_1^{-1}\sigma_2)[-1,-1,0]=(\sigma_1^{-1}\sigma_2)(\sigma_1^{-1}\sigma_2^{-1})(\sigma_2^{-1}\sigma_3^{-1}),\\
\nonumber \bar{u}^b &\rightarrow & (\sigma_1^{-1}\sigma_2)[-1,0,-1]=(\sigma_1^{-1}\sigma_2)(\sigma_3^{-1}\sigma_1^{-1})(\sigma_1^{-1}\sigma_2^{-1}),\\
 e^- &\rightarrow & (\sigma_1^{-1}\sigma_2)[-1,-1,-1]=(\sigma_1^{-1}\sigma_2)(\sigma_2^{-1}\sigma_3^{-1})(\sigma_3^{-1}\sigma_1^{-1})(\sigma_1^{-1}\sigma_2^{-1}).
\end{eqnarray}
Identifying\footnote{A footnote is in order to avoid potential confusion regarding the action of the conjugate $\dagger$ on braids. $\dagger:\sigma_i\mapsto \sigma^{-1}$ is simply the braid inverse which is an antiautomorpism. In the definitions of $\alpha_i$ and $\alpha_i^{\dagger}$ in Eqs. (\ref{alphadagger}) and (\ref{defalpha}), the order of braid generators is not reversed making the conjugation look like an automorphism. However, as shown in Eq.(\ref{example}), the vertical reflection corresponding to the braid inverse also reverses the action of the $\alpha_i$s from left to right to right to left. Restoring the left to right action then reverses the order again, giving the appearance of an automorphism.
\begin{eqnarray}
\left([0,0,0]\alpha_2\alpha_1\right) ^{\dagger}&=&(\alpha_1\alpha_2)^{\dagger}[0,0,0],\\
&=&[0,0,0]\alpha_2^{\dagger}\alpha_1^{\dagger}.
\end{eqnarray}}
\begin{eqnarray}\label{defalpha}
(\sigma_1^{-1}\sigma_2^{-1})=\alpha_1,\;\;(\sigma_3^{-1}\sigma_1^{-1})=\alpha_2,\;\;(\sigma_2^{-1}\sigma_3^{-1})=\alpha_3,\;\; \sigma_1^{-1}\sigma_2=\omega^{\dagger}\omega,
\end{eqnarray}
the antiparticles can this time be written as a right ideal as
\begin{eqnarray}
\nonumber S^d\equiv &{}&\\
\nonumber &{}&\;\; \bar{\nu}\omega^{\dagger}\omega +\\
\nonumber d^r\omega^{\dagger}\omega \alpha_1 &+& d^g\omega^{\dagger}\omega\alpha_2 + d^b\omega^{\dagger}\omega\alpha_3\\
\nonumber \bar{u}^r\omega^{\dagger}\omega\alpha_3\alpha_2 &+& \bar{u}^g\omega^{\dagger}\omega\alpha_1\alpha_3 + \bar{u}^b\omega^{\dagger}\omega\alpha_2\alpha_1\\
&+& e^{-}\omega^{\dagger}\omega\alpha_3\alpha_2\alpha_1.
\end{eqnarray}

Thus, the Helon braids correspond precisely to the basis states of one left and one right ideal of the complex chained octonions $\mathbb{C}\otimes\overleftarrow{\mathbb{O}}$.  The only exception here is the neutrino and anti-neutrino states which are identified differently in the two models. We here follow the identification as made by Furey, including the neutrino in the same minimal left ideal as the positively charged fermions. This is sensible because then all the fermions in a given ideal have the same sign for their isospin.

In the construction of minimal left ideals (reviewed in section (\ref{ideals})), $\omega$ and $\omega^{\dagger}$ are nilpotents defined as $\omega \equiv \alpha_1\alpha_2\alpha_3$ and $\omega^{\dagger}=\alpha_3^{\dagger}\alpha_2^{\dagger}\alpha_1^{\dagger}$. From these are constructed the idempotents $\omega\omega^{\dagger}$ and $\omega^{\dagger}\omega$. Using the identification of $\alpha_i$ and $\alpha_i^{\dagger}$ in terms of braid generators above one has
\begin{eqnarray}
\omega^{\dagger} &=& \alpha_3^{\dagger}\alpha_2^{\dagger}\alpha_1^{\dagger}=[0,0,0](\sigma_2\sigma_3)(\sigma_3\sigma_1)(\sigma_1\sigma_2)=[1,1,1].
\end{eqnarray}
Similarly, 
\begin{eqnarray}
\omega &=& \alpha_3\alpha_2\alpha_1=[0,0,0](\sigma_1^{-1}\sigma_3^{-1})(\sigma_3^{-1}\sigma_2^{-1})(\sigma_2^{-1}\sigma_1^{-1})=[-1,-1,-1].
\end{eqnarray}
Both $\omega$ and $\omega^{\dagger}$ defined in this way are pure braids\footnote{The definition of $\omega=\alpha_3\alpha_2\alpha_1$ differs by a minus sign from its definition of $\omega=\alpha_1\alpha_2\alpha_3$ in \cite{furey2016standard}. This, with the definition used here, both $\omega\omega^{\dagger}$ and $\omega^{\dagger}\omega$ pick up a physically irrelevant minus sign.}. A pure braid is one that does not permute the strands of the braids. They form a subgroup of a braid group and in this case $\omega$ and $\omega^{\dagger}$ are the center of $B_3^c$. Furthermore $\omega^{\dagger}\omega=\omega\omega^{\dagger}=[0,0,0]$, the untwisted unbraid (the identity). This is indeed an idempotent but indicates a conflict with the Helon model where the framed braid representing the neutrino (antineutrino) is not trivial, and is not an idempotent. In the Helon model, the weak interaction is represented topologically as the braid product therefore requiring nontrivial braiding. The symmetries of the minimal left ideals however are only the unbroken symmetries $SU(3)_c$ and $U(1)$. For these symmetries, the underlying braiding may be, and should be, trivial. Therefore this conflict is expected and does not indicate a contradiction.

Furthermore, $\alpha_i$ and $\alpha_i^{\dagger}$ commute with  $\omega^{\dagger}\omega=\omega\omega^{\dagger}$ and consequently the right ideal can be rewritten as the left ideal $S^d$ (repeated here for convenience)
\begin{eqnarray}
\nonumber S^d\equiv &{}&\\
\nonumber &{}&\;\;\bar{\nu} \omega^{\dagger}\omega+\\
\nonumber d^r\alpha_1\omega^{\dagger}\omega &+& d^g\alpha_2\omega^{\dagger}\omega + d^b\alpha_3\omega^{\dagger}\omega\\
\nonumber \bar{u}^r\alpha_3\alpha_2\omega^{\dagger}\omega &+& \bar{u}^g\alpha_1\alpha_3\omega^{\dagger}\omega + \bar{u}^b\alpha_2\alpha_1\omega^{\dagger}\omega\\
&+& e^{-}\alpha_3\alpha_2\alpha_1\omega^{\dagger}\omega.
\end{eqnarray}


\section{Discussion}

One of the most prominent challenges in theoretical physics today is understanding the theoretical origin of the SM gauge group along with why only some of the representations of these gauge groups are observed in Nature. Another is the unification of the SM with gravity. Recent attempts to use the NDAs, in particular the octonions to describe the symmetries of leptons and quarks has led to progress in the first challenge. The topological representation of leptons and quarks as framed braids has led to progress in the second challenge. This paper has shown that these two radically different models are connected. 

The Helon model represents a generation of fermions as simple braidings of three ribbons, connected at the top and bottom by nodes. The braiding and twisting of these ribbons are described mathematically by $B_3^c$ and $B_2$ respectively. Because these framed braids may be embedded into a braided ribbon network, a $q$-deformed generalisation of a spin-network, which are foundational in background independent approaches to quantum gravity, the Helon model makes feasible the development of a theory that unifies matter with quantum spacetime.

The generalized ideals of the complex quaternions $\mathbb{C}\otimes\mathbb{H}$ describe concisely all of the Lorentz representations of the SM, and the minimal left ideals of the complex octonions $\mathbb{C}\otimes\overleftarrow{\mathbb{O}}$ mirrors the behaviour of a single generation of leptons and quarks under the unbroken SM symmetries $SU(3)_c$ and $U(1)_{em}$. 

In the first part of this paper the Clifford Braiding Theorem of Kauffman and Lomonaco was used to show that each of the (hyper complex) normed division algebras admits a representation of a braid group. In particular, the braid groups $B_3^c$ and $B_2$ of the Helon model are precisely those that can be represented using $\mathbb{H}$ and $\mathbb{C}$ respectively.  

The minimal left ideals $\mathbb{C}\otimes\overleftarrow{\mathbb{O}}$ of are constructed using the Witt decomposition for $C\ell(6)$ and written in terms of a primitive idempotent defined as a product of nilpotent basis vectors of maximal totally isotropic subspaces. The main result of this paper is that by appropriately defining these basis vectors, which act as ladder operators, in terms of braid generators, the basis states of the minimal left ideals coincide with the framed braids found in the Helon model. 

It is the interchangeability of the twisting of ribbons and the braiding of them that makes connecting the two models possible. Although it was previously shown that any framed braid (composed of three three strands) may be written purely in terms of twists with trivial braiding, it was shown in this paper that Helon braids may also be written purely in terms of braiding in $B_3^c$ with trivial twisting. This is not true for framed braids in general, but it is precisely in writing the Helon braids in this pure braid form that it becomes possible to equate the basis vectors of the Witt decomposition of $C\ell(6)\cong\mathbb{C}\otimes\overleftarrow{\mathbb{O}}$ with certain products of braid generators in $B_3^c$.

The minimal left ideals are generated from the action of basis vectors on primitive idempotents representing the neutrino and antineutrino. These idempotents, when written as a braid correspond to the trivial braid. This indeed is an idempotent but indicates a conflict with the Helon model where the framed braid representing the neutrino (antineutrino) is not trivial, and is not an idempotent. However this is not alarming and rather should be expected. In the Helon model, the weak interaction is represented topologically as the braid product. The braid product is meaningless when the braiding is trivial since it will inevitable result in another trivial braid. Therefore a description of the weak force as a topological process requires nontrivial braiding. The symmetries of the minimal left ideals however are only the unbroken symmetries $SU(3)_c$ and $U(1)$. For these symmetries, the underlying braiding may be, and should be, trivial. This is indeed what was found here.  
    
This paper represents a first attempt at unifying two promising and interesting models describing the internal symmetries of leptons and quarks. The results obtained here are promising and connects these two model based on two radically different approaches. At the same time these results raise a number of questions left for future work.  

The Helon model is constructed out of two braid groups, $B_2$ and $B^c_3$, which can be represented using the complex numbers and quaternions. Yet, it is the minimal left ideals of the complex octonions, not the complex quaternions, that describe the unbroken symmetries of a generation of leptons and quarks. It remains to be shown how exactly the Helon braids as complex quaternions sit inside the complex octonions. It may be that the minimal left ideals pick out certain quaternionic subalgebras inside the octonions. 

What is lacking in the Helon model is any justification for the choice of braid group, other than $B_3^c$ being the smallest group that gives non-trivial framed braids. If one started with the NDAs then the braid groups at one's disposal would be dictated from these algebras. It is interesting that the braid groups represented by $\mathbb{C}$ and $\mathbb{H}$ are precisely those used in the Helon model. At the same time it begs the question of what the role of $B_7^c$ which finds a representation in the octonions might be. One may speculate that it might play a role in describing the color force, which in the Helon model is described in terms of 'braid stacking'. This remains to be investigated. 

Finally, braid groups are infinite. What mechanism is in place to select the finite number of braids that are physically relevant? It may be that using NDAs provides an answer to this question. Recall that the order of braid generators represented in terms of NDAs is always eight. This automatically leads to a finite set of possible braids. However, what exactly the set of possible braids is has not bee studied yet.
 

\appendix
\section{Braid group representations in terms of $C\ell^+(n,0)$ and $C\ell(0,n-1)$}
\begin{tabular}{|c|c|c|c|c|c|c|}
\hline 
 & $C\ell^+(2,0)$ & $C\ell(0,1)$  & $C\ell^+(3,0)$ & $C\ell(0,2)$ & $C\ell^+(7,0)$ & $C\ell(0,6)$\\ 
\hline 
$\sigma_1$ & $\frac{1}{\sqrt{2}}(1+e_{21})$ & $\frac{1}{\sqrt{2}}(1-e_{1})$ & $\frac{1}{\sqrt{2}}(1+e_{21})$ & $\frac{1}{\sqrt{2}}(1+e_{21})$& $\frac{1}{\sqrt{2}}(1+e_{21})$ & $\frac{1}{\sqrt{2}}(1+e_{21})$ \\
\hline 
$\sigma_2$ & & & $\frac{1}{\sqrt{2}}(1+e_{32})$ & $\frac{1}{\sqrt{2}}(1-e_{2})$ & $\frac{1}{\sqrt{2}}(1+e_{32})$ & $\frac{1}{\sqrt{2}}(1+e_{32})$\\ 
\hline 
$\sigma_3$ & & & $\frac{1}{\sqrt{2}}(1+e_{13})$ & $\frac{1}{\sqrt{2}}(1+e_{1})$ & $\frac{1}{\sqrt{2}}(1+e_{43})$ & $\frac{1}{\sqrt{2}}(1+e_{43})$ \\ 
\hline 
$\sigma_4$ & & & & &$\frac{1}{\sqrt{2}}(1+e_{54})$ & $\frac{1}{\sqrt{2}}(1+e_{54})$ \\ 
\hline 
$\sigma_5$ &  & & & &$\frac{1}{\sqrt{2}}(1+e_{65})$ & $\frac{1}{\sqrt{2}}(1+e_{65})$ \\ 
\hline 
$\sigma_6$ & & & & &$\frac{1}{\sqrt{2}}(1+e_{76})$ & $\frac{1}{\sqrt{2}}(1-e_{6})$ \\ 
\hline 
$\sigma_7$ & &  & & & $\frac{1}{\sqrt{2}}(1+e_{17})$ & $\frac{1}{\sqrt{2}}(1+e_{1})$ \\ 
\hline 
\end{tabular} 
\newline
\\

\acknowledgments

This work is supported in part National Natural Science Foundation of China grant RRSC0116. The author wishes to thank Cohl Furey, Louis Kauffman, Adam Gillard, and Benjamin Martin for insightful discussions and communications.


\bibliography{NielsReferences}  

\begin{thebibliography}{10}

\bibitem{Bilson-Thompson2005}
Sundance~O Bilson-Thompson.
\newblock A topological model of composite preons.
\newblock {\em arXiv preprint hep-ph/0503213}, 2005.

\bibitem{Bilson-Thompson2009}
Sundance Bilson-Thompson, Jonathan Hackett, and Louis~H Kauffman.
\newblock Particle topology, braids, and braided belts.
\newblock {\em Journal of Mathematical Physics}, 50(11):113505, 2009.

\bibitem{Bilson-Thompson2008}
Sundance Bilson-Thompson, Jonathan Hackett, Lou Kauffman, and Lee Smolin.
\newblock Particle identifications from symmetries of braided ribbon network
  invariants.
\newblock {\em arXiv preprint arXiv:0804.0037}, 2008.

\bibitem{Bilson-Thompson2012}
Sundance Bilson-Thompson.
\newblock {Braided Topology and the Emergence of Matter}.
\newblock In {\em Journal of Physics: Conference Series}, volume 360, page
  012056. IOP Publishing, 2012.

\bibitem{Bilson-Thompson2007}
Sundance~O Bilson-Thompson, Fotini Markopoulou, and Lee Smolin.
\newblock Quantum gravity and the standard model.
\newblock {\em Classical and Quantum Gravity}, 24(16):3975, 2007.

\bibitem{furey2016standard}
Cohl Furey.
\newblock Standard model physics from an algebra?
\newblock {\em arXiv preprint arXiv:1611.09182}, 2016.

\bibitem{gunaydin1973quark}
Murat G{\"u}naydin and Feza G{\"u}rsey.
\newblock Quark structure and octonions.
\newblock {\em Journal of Mathematical Physics}, 14(11):1651--1667, 1973.

\bibitem{gunaydin1974quark}
M~G{\"u}naydin and F~G{\"u}rsey.
\newblock Quark statistics and octonions.
\newblock {\em Physical Review D}, 9(12):3387, 1974.

\bibitem{dixon1990derivation}
Geoffrey Dixon.
\newblock Derivation of the standard model.
\newblock {\em Il Nuovo Cimento B (1971-1996)}, 105(3):349--364, 1990.

\bibitem{dixon2004division}
Geoffrey Dixon.
\newblock Division algebras: Family replication.
\newblock {\em Journal of mathematical physics}, 45(10):3878--3882, 2004.

\bibitem{dixon2010division}
Geoffrey~M Dixon.
\newblock Division algebras; spinors; idempotents; the algebraic structure of
  reality.
\newblock {\em arXiv preprint arXiv:1012.1304}, 2010.

\bibitem{manogue2010octonions}
Corinne~A Manogue and Tevian Dray.
\newblock Octonions, e6, and particle physics.
\newblock In {\em Journal of Physics: Conference Series}, volume 254, page
  012005. IOP Publishing, 2010.

\bibitem{furey2014generations}
Cohl Furey.
\newblock Generations: three prints, in colour.
\newblock {\em Journal of High Energy Physics}, 2014(10):1--11, 2014.

\bibitem{furey2018demonstration}
C~Furey.
\newblock A demonstration that electroweak theory can violate parity
  automatically (leptonic case).
\newblock {\em International Journal of Modern Physics A}, 33(04):1830005,
  2018.

\bibitem{stoica2017standard}
Ovidiu~Cristinel Stoica.
\newblock The standard model algebra.
\newblock {\em arXiv preprint arXiv:1702.04336}, 2017.

\bibitem{kauffman2016braiding}
Louis~H Kauffman and Samuel~J Lomonaco.
\newblock Braiding with majorana fermions.
\newblock In {\em SPIE Commercial+ Scientific Sensing and Imaging}, pages
  98730E--98730E. International Society for Optics and Photonics, 2016.

\bibitem{gresnigt2017quantum}
Niels~G Gresnigt.
\newblock Quantum groups and braid groups as fundamental symmetries.
\newblock {\em arXiv preprint arXiv:1711.09011}, 2017.

\bibitem{he2008c}
Song He and Yidun Wan.
\newblock C, p, and t of braid excitations in quantum gravity.
\newblock {\em Nuclear physics B}, 805(1):1--23, 2008.

\bibitem{smolin2008propagation}
Lee Smolin and Yidun Wan.
\newblock Propagation and interaction of chiral states in quantum gravity.
\newblock {\em Nuclear physics B}, 796(1):331--359, 2008.

\bibitem{baez2002octonions}
John Baez.
\newblock The octonions.
\newblock {\em Bulletin of the American Mathematical Society}, 39(2):145--205,
  2002.

\bibitem{doran2003gap}
C.~Doran and A.~N. Lasenby.
\newblock {\em {Geometric Algebra for Physicists}}.
\newblock Cambridge University Press New York, 2003.

\bibitem{hestenes1966sta}
D.~Hestenes.
\newblock {\em {Space-Time Algebra}}.
\newblock Gordon \& Breach Science Pub, 1966.

\bibitem{Gresnigt2009}
N.G. Gresnigt.
\newblock {\em {Relativistic Physics in the Clifford Algebra Cl(1,3)}}.
\newblock PhD thesis, University of Canterbury, 2009.

\bibitem{gresnigt2008}
N.~G. Gresnigt and P.~F. Renaud.
\newblock {On the Geometry of the Conformal Group in Spacetime}.
\newblock {\em Bulletin of the Belgian Mathematical Society-Simon Stevin 17
  (2), 193-200}, 2008.

\bibitem{gresnigt2007sph}
N.~G. Gresnigt, P.~F. Renaud, and P.~H. Butler.
\newblock {The Stabilized Poincare-Heisenberg Algebra: a Clifford Algebra
  Viewpoint}.
\newblock {\em Int. J. Mod. Phys. D}, 16(09):1519--1529, 2007.
\newblock (arXiv:hep-th/0611034).

\bibitem{lounesto2001caa}
P.~Lounesto.
\newblock {\em {Clifford Algebras and Spinors}}.
\newblock Cambridge University Press, 2001.

\end{thebibliography}
\bibliographystyle{unsrt}  



\end{document}